\documentclass[twocolumn]{aa}

\usepackage{graphicx}
\usepackage{txfonts}
\usepackage{color}

\begin{document}

\title{Missing baryons in shells around galaxy clusters}

\author{D. A. Prokhorov \inst{1,2} }

\offprints{D.A. Prokhorov \email{prokhoro@iap.fr}}

\institute{Institut d'Astrophysique de Paris, CNRS, UMR 7095,
Universit\'{e} Pierre et Marie Curie, 98bis Bd Arago, F-75014
Paris, France
            \and
            Moscow Institute of Physics and Technology,
            Institutskii lane, 141700 Moscow Region, Dolgoprudnii, Russia }

\date{Accepted . Received ; Draft printed: \today}

\authorrunning{D. A. Prokhorov}

\titlerunning{Missing baryons in shells around galaxy clusters}

\abstract {}{The cluster baryon fraction is estimated from the
CMB-scattering leptonic component of the intracluster medium
(ICM); however, the observed cluster baryon fraction is less than
the cosmic one. Understanding the origin of this discrepancy is
necessary for correctly describing  the structure of the ICM.} {We
estimate the baryonic mass in the outskirts of galaxy clusters
which is difficult to observe because of low electron temperature
and density in these regions.} {The time scale for the electrons
and protons to reach equipartition in the outskirts is longer than
the cluster age. Since thermal equilibrium is not achieved, a
significant fraction of the ICM baryons may be hidden in shells
around galaxy clusters. We derive the necessary condition on the
cluster mass for the concealment of missing baryons in an outer
baryon shell and show that this condition is fulfilled because
cluster masses are comparable to the estimated characteristic mass
$M=e^4/(m^3_{p} G^2)=1.3x10^{15}$ solar masses. The existence of
extreme-ultraviolet emission haloes around galaxy clusters is
predicted.}{}


\maketitle

\section{Introduction}

Clusters of galaxies are the largest virialized structures in the
Universe. They typically contain hundreds of galaxies, spread over
a region whose size is roughly 3 Mpc. Their total masses are about
$10^{15} M_{\bigodot}$. The space between galaxies in clusters is
filled with very hot $\sim 10^8$ K, low-density $\sim 10^{-3}$
cm$^{-3}$ gas (for a review, e.g. Sarazin 1986).

It is widely believed that clusters should be representative of
baryonic and nonbaryonic matter compositions, and thus that their
baryon fractions can be used to determine the average density of
the Universe in conjunction with Big Bang nucleosynthesis
(e.g. White et al. 1993).

Recent good quality observations of massive clusters with Chandra
and XMM-Newton put strong observational constraints on the gas
mass fraction $f_{b}$ in massive clusters at large radius (e.g.
Allen et al. 2004, Zhang et al. 2006). In many cases the new data
allow this fraction to be constrained out to the radius $r_{500}$
at which the density contrast is 500. The observable baryon
fraction is lower than the best fit to the WMAP 5-year result of
$f_{b}=0.17\pm0.01$ (Dunkley et al. 2008).

X-ray observations are not the only way to constrain the gas mass
fractions of clusters. The Sunyaev - Zeldovich (SZ) effect is also
being used for this purpose. LaRoque et al. (2006) estimated
$f_{b}$, in the radius at which the density contrast is 2500, by
fitting both BIMA/OVRO SZ effect data and Chandra X-ray data
simultaneously and fitting the SZ effect data alone. These results
are in excellent agreement with the X-ray derived results.

Measurements of the SZ effect in the CMB maps of WMAP data
(Afshordi et al. 2007), through study of a sample of 193 massive
galaxy clusters with observed X-ray temperature greater than 3
keV, indicate that a significant fraction, $35\%\pm8\%$, of
baryonic mass is missing from the hot ICM, and thus must cool to
form galaxies, intracluster stars, or an unknown cold phase of the
ICM. There does not seem to be enough mass in the form of stars or
cold gas in the cluster galaxies or intracluster space, signaling
the need for a still unknown baryonic component, or otherwise new
astrophysical processes in the ICM (Afshordi et al. 2007).

Much larger discrepancies with X-ray observations have been
reported in Lieu et al. (2006), who claim that the WMAP SZ signal
for ROSAT X-ray clusters is a factor of 4 weaker than the
expectation from X-ray brightness profiles.

However, in cosmological simulations the baryon fraction within
$r_{500}$ has converged to 90-95\% of the best fit to the WMAP
5-year data (Dunkley et al. 2008) in the case of non-radiative
simulations (e.g. Frenk et al. 1999, Kay et al. 2004, Crain et al.
2006). Cosmological cluster simulations that include gas cooling
and star formation also give an ICM mass fraction very similar to
the WMAP 5-year fraction (Kravtsov et al. 2005).

In a hierarchical Universe clusters form through accretion. As
fresh material joins the stationary cluster gas, it converts its
infall kinetic energy into heat through a shock. A shock front
moves outward as ambient gas accretes toward the cluster center
(see e.g. Takizawa \& Mineshige 1998). The accretion shock around
an X-ray cluster primarily heats the protons, since they carry
most of the kinetic energy of the infalling gas.  The electrons,
on the other hand, remain cold after the shock with their
temperature slowly rising as a result of their Coulomb collisions
with the hot ions. Thermal equilibrium will only be achieved if
the temperature equilibrium time between the protons and the
electrons is shorter than the age of the cluster (e.g. Fox \& Loeb
1997).

The Sunyaev-Zeldovich distortion of the CMB spectrum is
proportional to the electron temperature. Thus, the low gas
fractions inferred from the SZ effect observations in the CMB maps
of WMAP data (Lieu et al. 2006, Afshordi et al. 2007) can be
interpreted as evidence of a warm phase (with low electron
temperature $T_{e} \sim 0.1$ keV and high proton temperature) of
the ICM in the cluster outskirts, which would not contribute
appreciably to the SZ signal, while containing a significant
fraction of the ICM baryons.

The importance of cluster outskirt analysis is discussed in Sect.
2, where the baryon fraction profile is calculated in the case of
a simple isothermal beta model. The formation of baryonic shells
around clusters is considered within the framework of kinetic
theory in Sect. 3. A condition on the cluster mass for deviation
from thermal equilibrium is derived in Sect. 4. The influence of
the deviation from thermal equilibrium on the prediction for the
SZ decrement is considered in Sect. 5. A method for detecting
missing baryons by means of EUV astronomy is proposed in Sect. 6
and a discussion is presented in Sect. 7.

\section{Tracing the baryon fraction in the outskirts within the framework
of the hydrostatic theory}

The ICM emits energy mainly via thermal bremsstrahlung, which is
proportional to the density squared. Therefore external cluster
regions show less X-ray emission than the central ones, where the
density is much higher. This lower emission translates in lower
statistics for available X-ray observations. One possible way to
overcome the problem of low statistics in the outer regions is to
fit clusters with the isothermal beta model (Cavaliere \&
Fusco-Femiano 1976). The disadvantage of such a method is clear:
the extrapolation depends on the validity of the model in the
cluster outskirts. In accordance with the beta model, the electron
number density follows the spherical distribution
\begin{equation}
n(r)=n_{0}\left(1+\frac{r^2}{r^2_{c}}\right)^{-\frac{3\beta}{2}}
\label{beta}
\end{equation}
where $n_{0}$ is the central number density, $r_{c}$ is the core radius, $\beta$ is the slope parameter.

The hydrostatic equilibrium equation for a spherically symmetric
isothermal cluster (e.g. Sarazin 1986) can be written as:
\begin{equation}
kT dn = -\frac{G M(R) \mu m_{p} n(R) dR}{R^2}
\end{equation}
where $M(R)=\int^{R}_{0} 4\pi r^2 \psi(r)$ dr is the total mass
within a sphere of radius R, $\psi(r)$ is the mass density of all
the matter, $\mu$ the mean molecular weight, $m_{p}$ the proton
mass.

The baryonic fraction $f_{b}(r)=\rho(r)/\psi(r)$ is given by
\begin{equation}
f_{b}(r)=-\frac{4\pi\mu m_{p} G}{kT}\frac{\rho(r) r^2}{\left(r^2
\rho^{-1}(r)\cdot d\rho(r)/dr \right)^{\prime}}
\end{equation}
where $\rho(r) = m_{p} n(r)$ is the gas mass density, the
prime denotes a derivation of r.

Using Eq.(\ref{beta}), one can calculate the ratio
$f_{b}(r)/f_{b}(r_{c})$ for the isothermal beta model
\begin{equation}
F(r)=\frac{f_{b}(r)}{f_{b}(r_{c})} = 2^{3\beta/2}
\left(1+\frac{r^2}{r^{2}_{c}}\right)^{2-3\beta/2}\times\left(3+\frac{r^2}{r^{2}_{c}}\right)^{-1}.
\end{equation}

The maximum of the function $F(r)$ is at the point $a=r/r_{c}$:
\begin{equation}
a=\sqrt{\frac{9\beta-10}{2-3\beta}}, \ \ \ \ 2/3 <\beta <10/9.
\end{equation}

\begin{figure}[ht]
\centering
\includegraphics[angle=0, width=8cm]{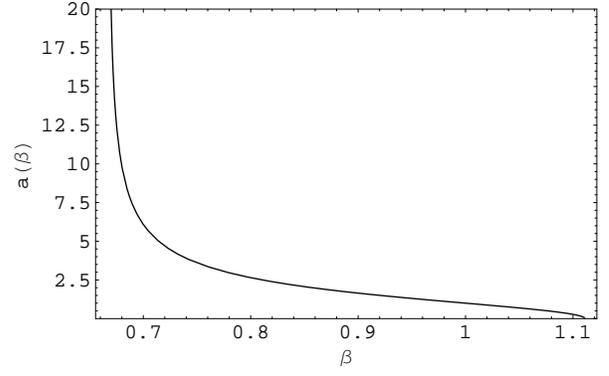}
\caption{Position of the maximum of the $f_{b}(r)/f_{b}(r_{c})$
ratio as a function of the slope parameter $\beta$.}
\end{figure}

The maximum value of $F(r)$ is given by
\begin{equation}
Fmax = \sqrt{(4-3\beta)^{4-3\beta} (3\beta-2)^{3\beta-2}}.
\end{equation}

\begin{figure}[ht]
\centering
\includegraphics[angle=0, width=8cm]{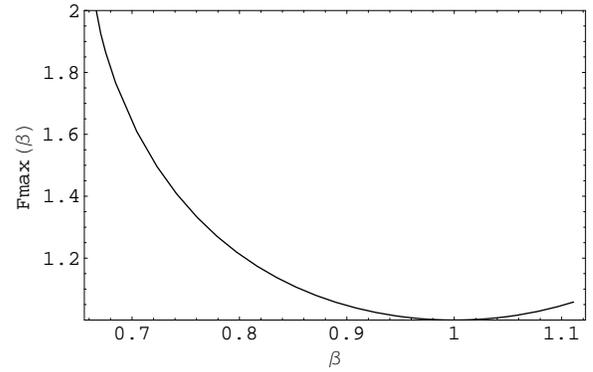}
\caption{Maximum value of the  $f_{b}(r)/f_{b}(r_{c})$ ratio as
a function of the slope parameter $\beta$.}
\end{figure}

The dependence $a(\beta)$ and the maximum value of $Fmax(\beta)$
are plotted in Figs. 1 and 2 respectively.

If $\beta<2/3$ then the baryon fraction $F(r)$ is an increasing
function of the radius. The average value of $\beta$ found by
Jones \& Forman (1984) from the X-ray distribution is
$\beta=0.65$, which implies that the gas is considerably more
extensively distributed than the mass in the cluster and the
baryon fraction in the outskirts is higher than in the central
region. Hence a study of the cluster outskirts is essential for
understanding the baryon component distribution in clusters.

The mass of the gas shell between the accretion shock radius
around a galaxy cluster and the virial radius $R_{vir}$ in the
context of the beta model (with $\beta<1$ and $r_{c}\ll R_{vir}$)
is given by integrating Eq.(1):
\begin{equation}
M^{gas}_{shell}\approx M^{gas}(R_{vir})\times
\left(\alpha^{3-3\beta} - 1\right)
\end{equation}
where $M^{gas}(R_{vir})$ is the gas mass inside the virial radius
sphere, and $\alpha$ is the ratio of the shock radius to the
virial radius.

Lieu et al. (2006) claim that the WMAP SZ signal for ROSAT X-ray
clusters is much weaker than the expectation from the X-ray
brightness profile. In order to compare with WMAP maps, they
extrapolate the isothermal beta model fits far beyond the region
that is fit by X-ray data. The reported discrepancy may be
evidence of a warm phase of the ICM in the cluster outskirts with
electron temperature $T_{e} \sim 0.1$ keV, which would not
contribute appreciably to the SZ signal, and the proton
temperature is of the order of the virial one (see Sect. 5).

\section {Formation of a baryonic shell within the framework of the kinetic
theory}

In accordance with the hierarchical clustering scenario, the
baryons are accreted along with the dark matter onto a galaxy
cluster. An accretion shock at the interface of the inner
hydrostatic gas with a cooler external medium is a long-standing
prediction from gravitationally driven models (e.g., see the model
of Bertschinger 1985).

The accretion shock converts the radial motion energy of the
upstream protons into random motion energy, below the shock.
Let us assume that the protons below the accretion shock
are essentially collisionless and have large orbital
eccentricities. Then they are able to accrete into the
X-ray cluster on the Coulomb time scale, since this is the time in
which they are able to dissipate their excess orbital angular
momentum (for a similar view of a quasar model, see Meszaros \&
Ostriker 1983). If this time scale is comparable with the cluster
age, then a considerable part of the protons can be located
between the accretion shock and the inner collisional plasma
region. The Coulomb mean-free-path of protons is larger than the
virial radius of the cluster at temperature $\gtrsim 2$ keV (Loeb
2007), therefore it is reasonable to choose the virial radius as
the inner collisional region border.

The Coulomb mean-free-path of a proton in a plasma of temperature
T is given by (Spitzer 1956):
\begin{equation}
l=\frac{3^{3/2} k^2_{B} T^2}{4 \pi^{1/2} n e^4 \ln{\Lambda}}
\label{lambda}
\end{equation}
where T is the proton temperature, n is the proton number density,
$\ln{\Lambda}$ is the Coulomb logarithm, e is the electron charge,
$k_{B}$ is the Boltzmann constant.

The proton random energy (or "temperature") is of the
order of the gravitational temperature.
\begin{equation}
T_{gr}=\frac{G M(R_{vir}) m_{p}}{k_{B} R_{vir}}
\label{Tgr}
\end{equation}
where $M(R_{vir})$ is the total mass inside the sphere of radius
$R_{vir}$, which equals (e.g. Eke et al. 1998)
\begin{equation}
R_{vir}=\left(\frac{3M(R_{vir})}{4\pi \Delta_{c}
\rho_{cr}}\right)^{1/3} \label{Rvir}
\end{equation}
where $\Delta_{c}$ is the density contrast for the formation of
the cluster.

The protons will be collisionless if the Coulomb mean-free path
$l$ is higher than the distance $L$, which the proton can cover
during the shock propagation time $t=\tau H^{-1}_{0}$ from the
virial radius to the present accretion shock radius:

\begin{equation}
\centering
l > L \label{comparison}
\end{equation}
where $H_{0}$ is the Hubble constant.

It is convenient to rewrite the distance $L$ by means of
Eqs. (\ref{Tgr}) and (\ref{Rvir}):
\begin{eqnarray}
&&L=\tau H^{-1}_{0}\sqrt{\frac{k_{B} T_{gr}}{m_{p}}} =\tau
R_{vir}\sqrt{\frac{\Delta_{c}}{2}}.
\end{eqnarray}

Using Eqs. (\ref{lambda}) and (\ref{comparison}), one can show
that the collisionless condition is equivalent to the inequality
\begin{equation}
n<\frac{3^{3/2} k^2_{B} T^2}{4\pi^{1/2} L e^4 \ln{\Lambda}}.
\end{equation}

The mass of the gas shell between the accretion shock and virial
radius is
\begin{equation}
M^{gas}_{shell} = \frac{4\pi}{3} R^3_{vir}
\left(\frac{R^3_{s}}{R^3_{vir}}-1\right) n m_{p}
\end{equation}
where $R_{s}$ is the present shock radius.

The overdensity in Bertschinger's collisionless model, given by
the dimensionless quantity $M(\lambda)/\lambda^3$, equals 178 at
$\lambda_{vir}\approx0.255$ (Fox \& Loeb 1997), where the
dimensionless parameter $\lambda=r/r_{ta}$ and $r_{ta}$ is the
turnaround radius. The shock occurs at fixed
$\lambda_{s}\approx0.347$ (for $\gamma=5/3$) in accordance with
the shocked accretion model of a collisional gas (Bertschinger
1985). Consequently, the ratio of the shock radius to the virial
radius $\alpha=R_{s}/R_{vir}$ equals 1.36.

One can rewrite the condition $l>L$ in terms of the shell mass
\begin{equation}
M^{gas}_{shell} < \frac{4\pi}{3} \left(\alpha^3-1\right) R^3_{vir}
m_{p}\frac{3^{3/2} k^2_{B} T^2}{4 \pi^{1/2} e^4 L \ln{\Lambda}}
\end{equation}
replacing $T$ by $T_{gr}$ from Eq. (\ref{Tgr}) one can get:
\begin{equation}
M^{gas}_{shell} < \tau^{-1} \sqrt{\frac{6\pi}{\Delta_{c}}}
\left(\alpha^3-1\right) m^3_{p} \frac{G^2 M^2(R_{vir})}{e^4
\ln{\Lambda}}.
\end{equation}
If the missing baryons are situated in the collisionless shell,
one can evaluate the value
$M^{gas}_{shell}/M^{gas}(R_{vir})=(f_{b}-f)/f$, where
$f=M^{gas}(R_{vir})/M(R_{vir})$ and $f_{b}$ is the best fit to the
WMAP 5-year result. Therefore
$M^{gas}_{shell}/M(R_{vir})=f_{b}-f$.

Thus a collisionless shell should exist if the total cluster mass
satisfies the necessary condition
\begin{equation}
M(R_{vir}) > \tau \sqrt{\frac{\Delta_{c}}{6\pi}} \frac{(f_{b}-f)
\ln{\Lambda}}{\alpha^3-1} \frac{e^4}{G^2 m^3_{p}}. \label{cond}
\end{equation}

The time evolution of the radius $r_{s}(t)$ of a shock surface
follows $r_{s}(t)\propto t^{8/9}$ (Bertschinger 1985). The
dimensionless shock propagation time from the virial radius to the
present accretion shock radius equals $\tau=1-\alpha^{-9/8}$,
assuming here that the cluster age is the Hubble time.

If $f_{b}-f=0.056$ (if 33\% of baryons are missing), $\alpha=1.36$
(the ratio of the shock radius to the virial radius in
Bertschinger's model) and $\tau=0.29$ (the shock propagation
time), then the total cluster mass must exceed the limit mass
value
\begin{equation}
M(R_{vir})>1.2\cdot 10^{15} M_{\bigodot}. \label{massa}
\label{Mlim}
\end{equation}

X-ray measurements tend to come from the inner parts of clusters.
One way to relate an observable quantity in the inner regions of a
cluster to a quantity in the outer regions is to use a simplified
model, such as that considered by Eke et al (1996), in which both
the X-ray gas and the total mass have the form of a singular
isothermal sphere in the virialized section of the cluster. In
this model, the mass within the sphere of the virial radius and
the temperature of the X-ray gas are related by
\begin{equation}
M(R_{vir})=1.7\cdot 10^{15} h^{-1}_{70} \left(\frac{T_{X}}{10^8
K}\right)^{3/2} M_{\bigodot}\label{Mvir}
\end{equation}
where $h_{70}$ is the Hubble constant in units of 70 km$\cdot$
s$^{-1}$ Mpc$^{-1}$.

The limit mass value $(1.2\cdot10^{15} M_{\bigodot})$ is
comparable with a typical cluster mass, therefore a fraction of
baryons can be trapped in the baryonic shell around clusters
within the framework of kinetic theory.

\section{Deviation from thermal equilibrium}

In Sections 2 and 3 using both hydrostatic and kinetic theories we
conclude that the fraction of cluster baryons located in the
outskirts may be significant. This raises the possibility that a
missing fraction of baryonic mass may be hidden in the cluster
outskirts where the electron temperature is much smaller than the
proton one, and therefore the contribution of these electrons to
the SZ signal is small.

The time scale $t_{eq}(p,e)$ for the electrons and protons to
reach equipartition $T_{e}=T_{p}$ is given by
$t_{eq}(p,e)\approx\sqrt{m_{p}/m_{e}}\cdot t_{eq}(p,p)$ (e.g. Sarazin 1986),
where the time scale $t_{eq}(p,p)=l/\sqrt{kT_{gr}/m_{p}}$ for
protons to equilibrate among themselves .

The electrons will be cold between the accretion shock and virial
radius if the time scale $t_{eq}(p,e)$ is longer than the shock
propagation time $t=\tau H^{-1}_{0}$ from the virial radius to the
present accretion shock radius, i.e. $t_{eq}(p,e)>\tau
H^{-1}_{0}$. This condition can be rewritten in the form
equivalent to Eq.(\ref{comparison}):

\begin{equation}
l>\sqrt{\frac{m_{e}}{m_{p}}}\cdot L.
\end{equation}

Assuming that the missing baryons are situated in the shell
between the accretion shock and virial radii and following the
same method as in Section 3, one can derive a necessary condition
for deviation from thermal equilibrium:

\begin{equation}
M(R_{vir}) > \tau \sqrt{\frac{m_{e}}{m_{p}}}
\sqrt{\frac{\Delta_{c}}{6\pi}} \frac{(f_{b}-f)
\ln{\Lambda}}{\alpha^3-1} \frac{e^4}{G^2 m^3_{p}}. \label{cond}
\end{equation}

If $f_{b}-f=0.056$ (if 33\% of baryons are missing), $\alpha=1.36$
(the ratio of the shock radius to the virial radius in
Bertschinger's model) and $\tau=0.29$ (the shock propagation
time), then the numerical value of this mass is
\begin{equation}
M(R_{vir})>2.8\cdot10^{13} M_{\bigodot}.
\end{equation}

The value of the limit mass $M(R_{vir})=2.8\cdot10^{13}
M_{\bigodot}$ is smaller than cluster masses for a sample that was
used by Afshordi et al. (2007) and therefore the missing baryons
may be hidden in shells around galaxy clusters. Note that even for
a value $\tau=1$ (the cluster age) the limit mass is still smaller
than cluster masses for the Afshordi et al. (2007) sample.

This explanation of the origin of the missing baryonic mass in the
clusters arises from a coincidence between a typical cluster mass
and a characteristic mass $(1.3\cdot10^{15} M_{\bigodot})$, which
is given by

\begin{equation}
M=\frac{e^4}{m^3_{p} G^2}.
\end{equation}

The characteristic mass depends only on fundamental physical
constants.

\section{Prediction for the SZ decrement}

The observed SZ effect with WMAP is much less than the predicted
decrement value given by the isothermal beta model (Lieu et al.
2006). Their study used a single-temperature model for the
intracluster medium, under the assumption that equipartition is
fully achieved. However, the outskirts of galaxy clusters actually
have a two-temperature structure, because the two temperatures,
i.e. those of electrons $T_{e}$ and protons $T_{p}$, remain
different (e.g. Fox \& Loeb 1997). To understand the discrepancy
between observed and predicted SZ decrements it is useful to study
the influence of the deviation from thermal equilibrium on the
prediction for the SZ decrement.

The thermal SZ effect (see Sunyaev \& Zeldovich 1980 for a review)
describes the inverse Compton scattering of CMB photons by the hot
gas in clusters. The resulting change in the (thermodynamic) CMB
temperature observed at frequency $\nu$ is given by
\begin{equation}
\Delta T_{\nu}=y j(x) T_{0}
\end{equation}
where $T_{0}=2.725$K is the unperturbed CMB temperature, $y$ is
the Comptonization parameter, $x$ is a dimensionless parameter
defined as $x=h\nu/kT_{0}$ and in the nonrelativistic regime
$(kT_{e}\ll mc^2)$
\begin{equation}
j(x)=x \frac{e^x + 1}{e^x - 1}-4.
\end{equation}
The Comptonization parameter is
\begin{equation}
y=\int dl \frac{kT_{e}}{m c^2} \sigma_{T} n
\end{equation}
where $T_{e}$ is the electron temperature, $\sigma_{T}$ is the
Thomson cross section, and the integral is over the distance along
the line-of-sight.

The total flux $S_{\nu}$ of a cluster (e.g. Refregier et al. 2000)
observed at frequency $\nu$ is related to the angular integral of
the thermodynamic temperature shift $\Delta T_{\nu}$ by
\begin{equation}
S_{\nu}=\frac{2 k^3 T^2_{0}}{h^2 c^2} q(x) \int d\Omega \Delta
T_{\nu}
\end{equation}
where $q(x)=x^4/[2 \sinh(x/2)]^2$. $\footnote{Note that this is
the correct formula, while there is a misprint in the expression
of q(x) in Eq.(19) by Refregier et al. (2000).}$ For the WMAP W
channel ($\nu=94$GHz, $x_{94}=1.65$): $q(x_{94})=2.18$ and the
full width at half maximum (FWHM) equals $10$ arcmin.

The postshock protons carry nearly all of the bulk kinetic energy.
After passing through a shock the electrons can gain thermal
energy via collisions with protons. The evolution of the electron
temperature is given by
\begin{equation}
\frac{dT_{e}}{dt}=\frac{T_{p}-T_{e}}{t_{eq}(p, e)}.\label{tep}
\end{equation}

The collision time scale $t_{eq}(p, e)$ (Spitzer 1958) equals
\begin{equation}
t_{eq}(p,e)=\frac{3m_{e}m_{p}}{8 \sqrt{2\pi} n e^4
\ln{\Lambda}}\left(\frac{k_{B} T_{e}}{m_{e}}+\frac{k_{B}
T_{p}}{m_{p}}\right)^{3/2}.\label{teq}
\end{equation}

\begin{figure}[ht]
\centering
\includegraphics[angle=0, width=8cm]{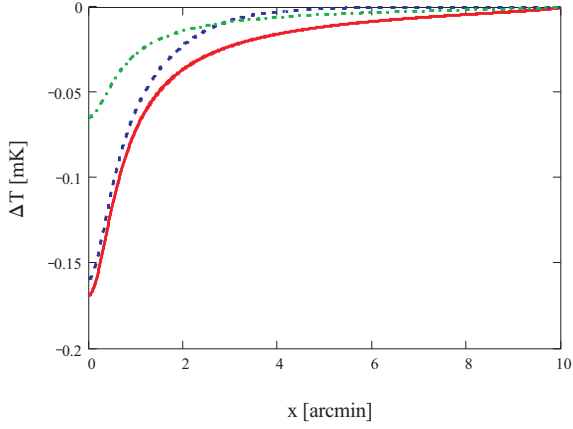}
\caption{SZ temperature decrements: for the isothermal beta model
(solid line), for the non-isothermal beta model (dashed line), and
for the isothermal beta model in which the value of the total SZ
flux corresponds to the non-isothermal model (dot-dashed line)}
\end{figure}

Using the dependence of the shock radius $r_{s}(t)\propto r^{8/9}$
from the work of Bertschinger (1985) and Eqs. (\ref{tep}),
(\ref{teq}) and assuming typical values of the central electron
temperature $k_{B}T_{0}=8$ keV, of the central density
$n_{0}=10^{-3}$ cm$^{-3}$, and of the slope parameter $\beta=2/3$,
we found the electron temperature profile and calculated both the
SZ temperature decrement and the total SZ flux $S_{\nu}$ within
$10'$ radius for the non-isothermal beta model. For a cluster with
a core radius $0.5'$ which corresponds to the mean core radius in
the sample of clusters by Bonamente et al. (2006), the SZ
decrements are plotted in Fig.3 for the isothermal and
non-isothermal beta models. The total SZ flux of a cluster in the
non-isothermal model is smaller by a factor of $2.6$ than those
computed with a single-temperature model. The SZ decrement for the
isothermal beta model which is scaled in such a way that the total
SZ flux corresponds to the non-isothermal model is also plotted in
Fig.3.

Although the total SZ flux for the non-isothermal beta model is
significantly lower than that for the isothermal model, attention
should be drawn to a paper of Bielby \& Shanks (2007), where they
truncated the isothermal beta model at the $2'$ radius and still
found a predicted SZ effect much larger than the level observed by
WMAP, for a sample of clusters (Bonamente et al. 2006). Note that
the approach of Afshordi et al. (2006) is different from that of
Bielby \& Shanks (2007), in the former the X-ray data are mainly
used to define a template to detect SZ decrements and then the SZ
data and the X-ray temperature data alone are used to establish
the gas densities.

\section{EUV emission from a baryonic shell}

Extreme ultraviolet (EUV) astronomy has produced substantial
results in a wide variety of fields (e.g. Bowyer et al. 2000). The
Extreme Ultraviolet Explorer (EUVE) has detected emission from a
few clusters of galaxies in the 70-200 eV energy range.

Since baryonic shells around clusters are a possible location of
missing cluster baryons and since the time scale of collisions
between baryons and electrons is longer than the Hubble time in
such shells, these cold electrons should be a source of
extreme-ultraviolet emission.

In this section the contribution to the total cluster EUV emission
from a baryonic shell is estimated for the Coma cluster. The Coma
cluster is a rich, hot ($T_{in}$=8 keV), nearby
($D=100h^{-1}_{70}$ Mpc) galaxy cluster, which is often considered
as a "standard cluster" due to the wealth of observational data
available for this object.

The spectral surface brightness along a particular line of sight
writes as
\begin{equation}
b(E)\sim\int n^2_{e}(\vec{r}) \Lambda(E, T_{e}) dl
\end{equation}
where $\Lambda (E, T_{e})$ is the spectral emissivity of the gas
at observed energy E, and $T_{e}$ is the electron temperature.

The spectral emissivity due to the electron-proton bremsstrahlung
rate is given by
\begin{equation}
\Lambda(E, T_{e})\sim\frac{1}{\sqrt{T_{e}}} \exp\left(-
\frac{E}{k_{B} T_{e}}\right).
\end{equation}

For the sake of clarity, we propose that the hot gas is located
inside the virial sphere. The electrons remain cold inside the
baryonic shell, because the time scale of their collisions with
the hot protons is longer than the Hubble time. The temperature of
these electrons is almost the same as the preshock temperature, a
plausible value is $T_{ext}=3\cdot10^6$ K.

The gas density is given by the isothermal beta model (Eq.~1)
within the virial sphere, and the gas density inside the shell,
which corresponds to missing baryons in the galaxy cluster (if
33\% of baryons are missing), equals
\begin{equation}
n_{shell}=0.5 \int^{R_{vir}}_{0} n(r) 4\pi r^2 dr \cdot
\left(\frac{4\pi}{3}R^3_{s} - \frac{4\pi}{3}R^3_{vir}\right)^{-1}
\end{equation}
where the center number density $n_{0}=3.4\cdot10^{-3}$ cm$^{-3}$,
the core radius $r_{c}=290$ kpc, $\beta=0.75$, the virial radius
$R_{vir}=2.8$ Mpc for a Coma-like cluster ($H_{0}=70$ km$\cdot$
s$^{-1}$ Mpc$^{-1}$ is adopted).

The spectral surface brightness from the hot gas is

\begin{equation}
b_{h}(\theta) \sim \int^{\Phi_{max}(\theta)}_{-\Phi_{max}(\theta)}
n^2_{0}\cdot \left(1+\frac{D^2\sin^2{\theta}}{r^2_{c}
\cos^2{\Phi}}\right)^{-3\beta}\frac{e^{-E/T_{in}}}{\sqrt{T_{in}}}
\frac{D\sin{\theta}}{\cos^2{\Phi}}
 d\Phi
\end{equation}
where $\theta$ is the angle between the center of the cluster and
the direction of interest $\Phi_{max}$ is defined by
\begin{equation}
\Phi_{max}=arctg\left(\frac{\sqrt{R^2_{vir}-D^2\sin^2{\theta}}}{D\sin{\theta}}\right)
\end{equation}
where D is the distance of the cluster.

The spectral surface brightness from the cold electrons is given
by

if $\theta<R_{vir}/D$:
\begin{equation}
b_{c}(\theta) \sim 2 \int^{\Phi_{max
2}(\theta)}_{\Phi_{max}(\theta)} n^2_{shell}
\frac{e^{-E/T_{ext}}}{\sqrt{T_{ext}}}
\frac{D\sin{\theta}}{\cos^2{\Phi}}
 d\Phi
\end{equation}

if $\theta\geq R_{vir}/D$:
\begin{equation}
b_{c}(\theta) \sim \int^{\Phi_{max 2}(\theta)}_{-\Phi_{max
2}(\theta)} n^2_{shell} \frac{e^{-E/T_{ext}}}{\sqrt{T_{ext}}}
\frac{D\sin{\theta}}{\cos^2{\Phi}}
 d\Phi
\end{equation}
where $\Phi_{max 2}$ is

\begin{equation}
\Phi_{max 2}
=arctg\left(\frac{\sqrt{R^2_{s}-D^2\sin^2{\theta}}}{D\sin{\theta}}\right).
\end{equation}

The normalized spectral surface brightness writes as
\begin{equation}
B(\theta)=\frac{b_{h}(\theta)+b_{c}(\theta)}{b_{h}(0)+b_{c}(0)}.
\end{equation}

For the Coma cluster, the normalized spectral brightness at observed
energy 0.1 keV is plotted in Fig. 3, where $1^{o}$ is 1.7~Mpc. Fig.3
shows the presence of an extreme-ultraviolet emission halo around the
galaxy cluster without the X-ray emission from it, because only the cold
electrons are situated in the baryonic shell.

\begin{figure}[ht]
\centering
\includegraphics[angle=0, width=8cm]{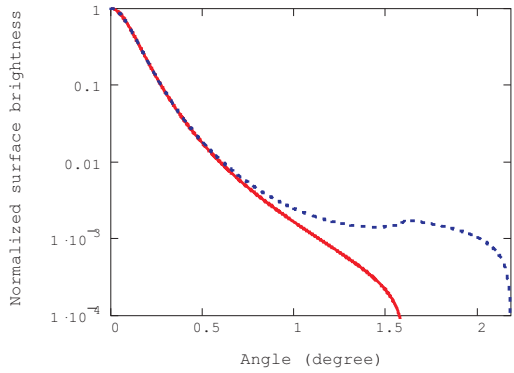}
\caption{Normalized surface brightness for the Coma cluster: the
hot gas (solid line), the hot gas + the baryonic shell (dashed
line).}
\end{figure}

\section{Discussion and conclusions}

Cluster gas mass fractions as inferred from X-ray observations
have been used as a probe of the universal ratio of baryon to
total mass densities $\Omega_{B}/\Omega_{M}$ (e.g. White et al.
1993, David et al. 1995, Evrard 1997, Roussel et al 2000, Allen et
al. 2002, Ettori 2003, Allen et al. 2004). But the result of
recent measurements of SZ effect and X-ray studies is that a
significant fraction $35\%\pm8\%$ of the baryons is missing from
the ICM (Ettori 2003, LaRoque et al. 2006, Vikhlinin et al. 2006,
Afshordi et al. 2007, McCarthy et al. 2007). A possible
explanation is that a significant fraction of baryonic mass may be
hidden in a form that is difficult to detect with ordinary
observational means.

The problem is unlikely to be associated with the conventional
cluster mass estimates at least within the Abell radius, which
rely upon the hydrostatic equilibrium hypothesis for the dynamical
state of clusters in the computation of their total masses (e.g.
Wu \& Xue 1999). This point has been justified by the excellent
agreement among the X-ray, optical and weak lensing determined
cluster masses on Abell radius scales (e.g. Allen 1998, Wu et al.
1998). However how accurate the conventional cluster mass
estimates will be in the outskirts of clusters where hydrostatic
equilibrium becomes questionable remains unclear.

We conclude within the framework of hydrostatic theory that the
gas is considerably more extensively distributed than the mass in
the cluster. Therefore a significant fraction of the ICM baryons
may be located in shells around galaxy clusters.

In this paper the formation of baryonic shells around clusters is
considered within the framework of the kinetic theory. Baryonic
shells can be formed in this way for galaxy clusters with masses
($M>1.2\cdot10^{15} M_{\bigodot}$).

We propose that the low baryon fraction inferred from the SZ
effect observations in the CMB maps of WMAP data (Afshordi et al.
2007) can be explained by the deviation from thermal equilibrium
in the cluster outskirts. The electrons and protons do not reach
equipartition there. This warm phase (with electron temperature
$T_{e}\sim0.1$ keV and with high proton temperature) of the ICM in
the cluster outskirts does not contribute appreciably to the SZ
signal. The necessary condition on the cluster mass for the
concealment of the cluster missing baryons in a baryon shell is
fulfilled because cluster masses are comparable to the value of
the characteristic mass $M=e^4/(m^3_{p}G^2)=1.3\cdot10^{15}
M_{\bigodot}$.

We considered a non-isothermal beta model for the ionized gas
distribution of a cluster and showed that the total SZ flux of a
cluster differs by a factor of $2.6$ from that computed with a
single-temperature beta model.

The cold electrons in shells around galaxy clusters should be a
source of extreme-ultraviolet emission. Observations of cluster
outskirts can shed light on a location of missing baryons.

Other possibilities that have been proposed to solve the cluster
missing baryon problem are: diffuse intracluster light, which would be
challenging to detect observationally, but could potentially hide
significant amounts of stellar mass in the vast intracluster space
(Lin \& Mohr 2004, Gonzales et al. 2005, Zibetti et al. 2005,
Monaco et al 2006); hiding baryons in the form of cold and compact
dark baryonic clouds from local cooling instabilities within the ICM
(Dwarakanath et al. 1994); evaporation of baryons out of the
virial radius of the cluster (Loeb, 2007); the correct value of
$\Omega_{M}$ is higher than the best-fit WMAP value (McCarthy et
al. 2007).

\begin{acknowledgements}
I am grateful to Florence Durret and Roya Colin Mohayaee for
valuable discussions and to the referee for constructive comments.
\end{acknowledgements}

\end{document}